\documentclass[lettersize,journal]{IEEEtran}
\usepackage{amsmath,amsfonts}
\usepackage{algorithmic}
\usepackage{algorithm}
\usepackage{array}
\usepackage[caption=false,font=normalsize,labelfont=sf,textfont=sf]{subfig}
\usepackage{textcomp}
\usepackage[table]{xcolor}

\usepackage{stfloats}
\usepackage{url}
\usepackage{color}
\usepackage{verbatim}
\usepackage{graphicx}
\usepackage{cite}
\usepackage{threeparttable}
\usepackage{amsmath,graphicx}
\usepackage{bbm}
\usepackage{tabularx}
\usepackage{booktabs}
\usepackage{textcomp}
\usepackage{multirow}
\usepackage{cite}
\usepackage{tikz}
\usepackage{threeparttable}
\usepackage{adjustbox}
\usepackage{diagbox}
\usepackage{setspace}
\usepackage{CJKutf8}

\usepackage{hyperref}

\newcommand{\mycc}{\cellcolor[HTML]{f2f2f2}}
\begin{document}
\title{Zero-Shot Audio Captioning Using\\ Soft and Hard Prompts}

\author{Yiming Zhang, Xuenan Xu, Ruoyi Du, Haohe Liu, Yuan Dong, Zheng-Hua Tan, \IEEEmembership{Senior Member, IEEE}, \\Wenwu Wang, \IEEEmembership{Senior Member, IEEE}, Zhanyu Ma, \IEEEmembership{Senior Member, IEEE}
\thanks{Y. Zhang, R. Du, D. Yuan, and Z. Ma are with the Pattern Recognition and Intelligent System Laboratory, School of Artificial Intelligence, Beijing University of Posts and Telecommunications, Beijing 100876,
China. E-mail: \{zhangyiming, duruoyi, mazhanyu, yuandong\}@bupt.edu.cn.}
\thanks{X. Xu is with the Department of Computer Science and Engineering,
Shanghai Jiao Tong University, Shanghai, 200240, China. Email: wsntxxn@sjtu.edu.cn.}
\thanks{Z.-H. Tan is with the Department of Electronic Systems, Aalborg University, Aalborg 9220, Denmark. E-mail: 
zt@es.aau.dk.}
\thanks{H. Liu, W. Wang is with the Centre for Vision, Speech and Signal Processing, University of Surrey, Guildford, GU2 7XH, United Kingdom. E-mail: \{haohe.liu, w.wang\}@surrey.ac.uk.}

\thanks{(Corresponding author: Zhanyu Ma)}}

\maketitle
\begin{abstract}
In traditional audio captioning methods, a model is usually trained in a fully supervised manner using a human-annotated dataset containing audio-text pairs and then evaluated on the test sets from the same dataset.
Such methods have two limitations. First, these methods are often data-hungry and require time-consuming and expensive human annotations to obtain audio-text pairs. Second, these models often suffer from performance degradation in cross-domain scenarios, i.e., when the input audio comes from a different domain than the training set, which, however, has received little attention. 
We propose an effective audio captioning method based on the contrastive language-audio pre-training (CLAP) model to address these issues. Our proposed method requires only textual data for training, enabling the model to generate text from the textual feature in the cross-modal semantic space.
In the inference stage, the model generates the descriptive text for the given audio from the audio feature by leveraging the audio-text alignment from CLAP.
We devise two strategies to mitigate the discrepancy between text and audio embeddings: a mixed-augmentation-based soft prompt and a retrieval-based acoustic-aware hard prompt.
These approaches are designed to enhance the generalization performance of our proposed model, facilitating the model to generate captions more robustly and accurately.
Extensive experiments on AudioCaps and Clotho benchmarks show the effectiveness of our proposed method, which outperforms other zero-shot audio captioning approaches for in-domain scenarios and outperforms the compared methods for cross-domain scenarios, underscoring the generalization ability of our method.

\end{abstract}
\begin{IEEEkeywords} Audio captioning, zero-shot, contrastive language-audio pre-training, prompt engineering 
\end{IEEEkeywords}
\section{Introduction}

\IEEEPARstart{A}{udio} captioning is a sophisticated audio-to-text cross-modal translation task where a model is built to analyse the contents of an audio clip and articulate it using natural language~\cite{drossos2017automated,drossos2020clotho,kim2019audiocaps,lipping2019crowdsourcing,xu2023beyond}.
The generated captions encompass not only basic descriptions of sound events and scenes but also high-level semantic information, such as the relationships among events and physical properties of sounds.
This complex integration enables a deeper contextual interpretation of audio data.
Audio captioning holds significant potential applications across diverse fields, including assistance for the hard of hearing, subtitles for television programs, and audio-text cross-modal retrieval.  

Recent advancements in audio captioning have significantly elevated the state-of-the-art.
However, most existing methods rely on fully supervised training, employing an audio-encoder
coupled with a language-decoder framework. Therefore these approaches are data-hungry and rely on large amounts of human-annotated audio description data for training.
Yet, data scarcity is a substantial challenge for audio captioning.
The predominant audio captioning benchmark datasets, Clotho~\cite{drossos2020clotho} and AudioCaps~\cite{kim2019audiocaps} contain only $19$k and $49$k audio-caption pairs in their training sets, respectively.
These numbers pale in comparison to the vast datasets available for visual captioning {(\textit{e.g.}, about $414$K paired data in the COCO Caption dataset~\cite{chen2015microsoft}).}

\begin{figure}
\centerline{\includegraphics[width=9cm]{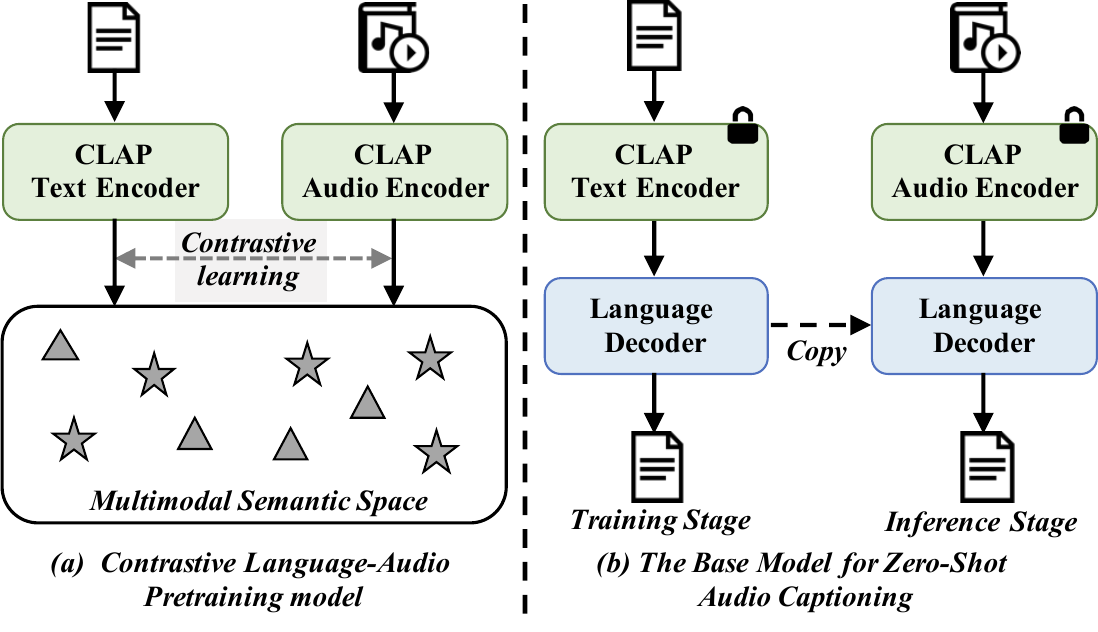}}

\caption{(a) The structure of the CLAP model. Through contrast learning, CLAP maps the audio and text into the same semantic space. Grey triangles and pentagons represent audio and text embeddings, respectively. (b) The structure of the base zero-shot audio captioning model, where a language decoder is trained for text reconstruction using text data based on the CLAP text encoder. The CLAP audio encoder is combined with the language decoder to generate captions during inference.}
\vspace{-10pt}
\label{fig:basemeth}
\end{figure}
The underlying reason for this predicament lies in the complex and costly process of annotating audio captioning datasets.
{ The audio is time-series data and has ambiguous properties, necessitating annotators to thoroughly attend it and conduct complex analyses to ensure accurate descriptions~\cite{zhang2023actual}.}
To alleviate the challenges of data annotation, researchers~\cite{kim2019audiocaps,martin2021diversity,wu2023large,mei2023wavcaps} have employed supplementary information (\textit{e.g.}, visual cues, and audio category details) or data augmentation techniques (\textit{e.g.}, text mixing, and large language model (LLM)).
While these approaches expand the dataset scale, they introduce biases and noise that potentially impact dataset quality~\cite{martin2021diversity}.

In addition, most existing studies typically evaluate the model performance solely in in-domain scenarios, where the training and test sets come from the same source. 

Accordingly, cross-domain scenarios where the training and test sets come from different sources receive little attention, although they happen more commonly in real-world applications.
These existing methods are often trained using limited in-domain data, which can result in model overfitting. Consequently, they can suffer from significant performance degradation in cross-domain scenarios and fail to describe out-of-domain audio clips accurately.

To address this issue, we propose a zero-shot audio captioning method to alleviate the reliance of the model on audio-text paired data and improve its generalization performance.
We adopt the contrastive language-audio pre-training model (CLAP)~\cite{mei2023wavcaps}, which constructs an implicit audio-text multimodal semantic space based on contrastive learning, as the backbone of the encoder which is shown in Fig.~\ref{fig:basemeth}.(a).
We only use textual data for training, making the training possible in scenarios where audio-text pairs are missing.
Captions can be generated by replacing the CLAP text encoder with the CLAP audio encoder during inference. 
However, the CLAP model struggles to construct a well-aligned multimodal semantic space and still exhibits a \textit{Modality Gap}~\cite{liang2022mind}, which renders simply replacing the encoder during the inference stage ineffective.
To bridge the modality gap of the CLAP model, We devise a mixed-augmentation strategy, which contains instance replacement and embedding augmentation, to improve the robustness and performance of the proposed model.
Meanwhile, to further improve the generalization performance of the model, we introduce the retrieval-based acoustic-aware prompt strategy,  which provides explicit acoustic information.

Overall, our main contributions are as follows.
\begin{itemize}
\item [1)] Focusing on zero-shot audio captioning,
we propose a simple yet effective method that uses only textual data to train the model and then generate captions for given audio clips during inference.
\item [2)] We devise the mixed-augmentation-based soft prompt to bridge the gap between the training and inference and introduce the acoustic-aware hard prompt to enhance the generalization of the proposed model.
\item [3)] Through extensive experimentation, we demonstrate the superior performance of our proposed method as compared with previous zero-shot audio captioning methods for in-domain scenarios, and fully supervised and zero-shot audio captioning methods for cross-domain scenarios.
\end{itemize}

\section{Related Work}
In this section, we first give a brief overview of CLAP, whose multimodal semantic space provides the foundation of our proposed method. 
Then, we introduce traditional fully supervised audio captioning methods and recent zero-shot audio captioning methods.

\subsection{Contrastive Language-Audio Pre-training (CLAP)}
CLAP~\cite{mei2023wavcaps,wu2023large,10095889,elizalde2024natural} utilizes contrastive learning to pre-train language-audio models, which map both audio and text into the same semantic space on large-scale audio-text pairs.
CLAP contains two encoders: an audio encoder and a text encoder. 
The audio encoder $f_{clap}^{Audio}(\cdot)$ often uses well-performed audio classification models, which can be convolution neural networks~\cite{kong2020panns} or Transformers~\cite{chen2022hts}, as the backbone.

The text encoder $f_{clap}^{Text}(\cdot)$ is usually a pre-trained masked language model (\textit{e.g.}, BERT~\cite{devlin2018bert}, RoBERTa~\cite{liu2019roberta}).
CLAP utilizes noisy pairwise data for training based on the InfoNCE loss~\cite{oord2018representation}, learning the alignment between text and audio embeddings in a multimodal semantic space.

In this work, we use CLAP text encoder $f_{clap}^{Text}(\cdot)$ for text reconstruction in the training stage.
In the inference stage, $f_{clap}^{Text}(\cdot)$ is replaced with the audio encoder $f_{clap}^{Audio}(\cdot)$ to generate the descriptive text for a given audio.

\subsection{Fully Supervised Audio Captioning}
With the success of DCASE challenges~\cite{lipping2019crowdsourcing}, fully supervised audio captioning has seen significant advancements.
Most research on audio captioning utilizes an audio encoder-language decoder framework trained on human-annotated audio-text paired data.
These studies employed the audio encoder to extract embeddings of the input audio clip $A$, which are then fed into the language decoder to generate corresponding descriptive caption $T$.
Mei \emph{et al.}~\cite{mei2021audio} proposed a full Transformer-based audio captioning method to improve the capability of modelling global and fine-grained temporal information.
Ye \emph{et al.} ~\cite{ye2021improving} proposed a fully supervised audio captioning model based on the multi-modal attention module, which utilizes acoustic and semantic information to generate captions.
Xu \emph{et al.}~\cite{xu2022sjtu} pre-trained the audio encoder on text-audio retrieval tasks, enhancing the representation capability of the audio encoder for audio captioning.
Kim \emph{et al.}~\cite{kim2023prefix} used a pre-trained language model (GPT-2) as the decoder to ensure text generation capability, with global and temporal information from the input audio as the prefix to guide the output of the decoder.
Koh \emph{et al.}~\cite{koh2022automated}  introduced the reconstruction latent space similarity regularisation to regulate model training in audio captioning.
Zhang \emph{et al.}~\cite{zhang2023actual} proposed a two-stage audio captioning approach to mitigate the effects of semantic disparity among the audio captions by incorporating feature space regularisation and improving the accuracy of the model-generated description text.
Ghosh \emph{et al.}~\cite{ghosh2024recap} proposed a retrieval-augmented audio captioning method that uses the CLAP encoder to retrieve captions similar to the input audio from the external database and then the retrieved captions are used as extra guidance for the decoder to generate descriptive text.

However, the high cost of collecting audio-text paired data has limited the applicability of these methods.
Therefore, reducing the dependency of audio captioning models on paired data has emerged as a prominent research focus in audio captioning.
\begin{figure*}
\centerline{\includegraphics[width=18cm]{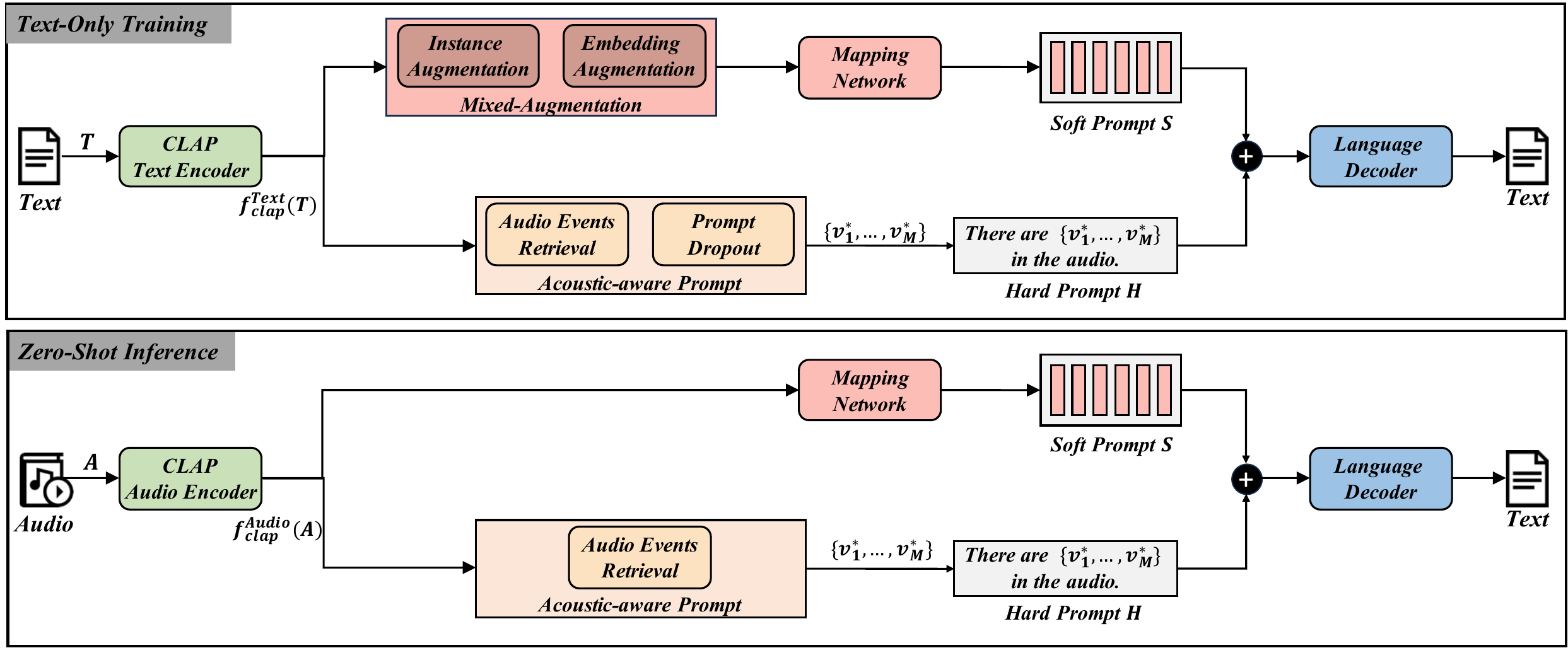}}

\caption{The overall architecture of our proposed method. Specifically, in the training stage, { we reconstruct the input text based on acoustic-aware prompts and soft prompts with only textual data}, so training does not require any paired data. During inference, we replace the CLAP text encoder $f_{clap}^{Text}(\cdot)$ with the CLAP audio encoder $f_{clap}^{Audio}(\cdot)$ to generate the descriptive text of the input audio.}
\vspace{-10pt}
\label{fig:method}
\end{figure*}
\subsection{Zero-Shot Audio Captioning}
To further reduce the cost of paired data collection, zero-shot audio captioning aims to generate audio captions without prior training for this task~\cite{shaharabany2023zero}.
Audio Flamingo~\cite{kong2024audio} used a large-scale weakly aligned audio-text pair dataset to train the audio language model and evaluated the model on the Audiocaps benchmark without fine-tuning.
Some works conducted zero-shot audio captioning by combining pre-trained audio-text models and large language models.
{ We categorize these studies into decoder-guided and encoder-guided methods based on where the acoustic information was introduced.
In the decoder-guided methods, the acoustic information is injected after the word probabilities are predicted by the language decoder.} Shaharabany \emph{et al.}~\cite{shaharabany2023zero} designed a classifier-guided zero-shot approach in which only audio data is used to optimize the hidden states of the language model to generate descriptive text with audibility.
Salewski \emph{et al.}~\cite{salewski2023zero} proposed a similar approach, where audio data is not used to optimize the hidden states, but to reweight the probability of output words.
However, decoder-guided approaches usually achieve poor performance, cannot achieve satisfactory zero-shot capability and the generated captions fail to describe the audio content accurately.
{ Compared to decoder-guided methods, encoder-guided methods rely more on the multimodal modelling capabilities provided by a pre-trained text-audio model (\textit{e.g.} CLAP), and the acoustic information is taken as input to the language decoder. }To mitigate the \textit{Modality Gap}~\cite{liang2022mind}, Deshmukh \emph{et al.}~\cite{deshmukh2024training} injected random variables into the text-only training. In contrast, Kouzelis  \emph{et al.}~\cite{kouzelis2023weakly} mapped the input CLAP audio embeddings to text embeddings in the inference stage to generate descriptive text. Although these encoder-guided methods can perform better in in-domain situations, they often overlook cross-domain scenarios.

In comparison to these methods, we propose an encoder-guided zero-shot audio captioning method, in which the mixed augmentation strategy is integrated to alleviate the problem of \textit{Modality Gap} and the auditory-aware prompt strategy is used to further enhance the accuracy of the generation by providing explicitly the external acoustic knowledge.

\section{Proposed Method}
{In this work, we propose a zero-shot audio captioning method to alleviate the reliance of the model on audio-text paired data in traditional fully supervised audio captioning methods.
The overall architecture of our proposed method is illustrated in Fig.~\ref{fig:method}. In the training stage, we use the CLAP text encoder to extract the embedding of the input text, and then the soft prompt and acoustic-aware hard prompt are fed to the language decoder to reconstruct the given text. In the inference stage, we shift from text-to-text generation to audio-to-text generation by replacing the CLAP text encoder with the CLAP audio encoder.} 

\subsection{The Soft Prompt based on Mixed-augmentations}
An intuitive method for the zero-shot audio captioning task is shown in Fig~\ref{fig:basemeth} (b). During training, for a given input text $T$ from the corpus $\mathcal{T}$, the language decoder is trained using the CLAP model to reconstruct the input text. During inference, only the text encoder needs to be replaced with an audio encoder to generate descriptive text for the input audio clip. However, due to the modality gap in the CLAP model, the model trained in this way can be limited in its generalization ability. To address this issue, we employ a mixed-augmentations strategy, which includes instance replacement and embedding augmentation, to enable the model to learn more robust latent representations.

\vspace{2pt}
\noindent \textbf {\textit{Instance Replacement:}}
First, we retrieve $N$ captions in the text corpus $\mathcal{T}$ that are semantically similar to the input text $T$ as a semantic candidate set $\mathcal{C}_N$:
\begin{equation}
\mathcal{C}_N = \left\{ \underset{T^{*}_n \in \mathcal{T}}{\mathrm{argmax}_{N}}\frac{f_{{clap}}^{{Text}}(T) \cdot f_{{clap}}^{{Text}}(T^{*}_n)}{\Vert f_{{clap}}^{{Text}}(T) \Vert\cdot\Vert f_{{clap}}^{{Text}}(T^{*}_n) \Vert}\right\},
\end{equation}
where $\mathrm{argmax}_{N}$ select text embeddings with top-$N$ highest similarities,  $f_{clap}^{Text}(T)$ is the CLAP text embedding of the input text $T$, $\Vert \cdot \Vert$ represents the norm of the embedding, $T^{*}_n$ is the $n$-th candidate text, and $n \leq N$.

Then, $f_{{clap}}^{{Text}}(T^{*}_n)$ is randomly selected from the candidate text embeddings set $\mathcal{C}_N$ to replace the original text embedding $f_{{clap}}^{{Text}}(T)$.

\vspace{3pt}
\noindent \textit{\textbf{Embedding Augmentation:}} To encourage the model to learn more robust latent representations, we insert a Gaussian noise $\epsilon \sim \mathcal{N}(0$, $\sigma)$ into the candidate text embedding $f_{{clap}}^{{Text}}(T^{*}_n)$ to obtain the noisy text embedding $f_{{clap}}^{{Text}}(T^{*}_n) + \epsilon$, where $\sigma$ is the standard deviation. 

Then, the noisy text embedding is fed into the mapping network $\mathcal{M}(\cdot)$ to get the soft prompt $S$ for the language decoder,
\begin{equation}
S= \mathcal{M}\left(f_{{clap}}^{{Text}}(T^{*}_n) + \epsilon \right),
\end{equation}
where $S = \left\{s_1,\dots,s_K \right\}$, $s_k$ is the $k$-th soft prompt embedding, and $K$ is the total length of the soft prompts $S$.

\subsection{Acoustic-aware Prompt based on Retrieval}
Acoustic labels are well-defined representations of the content and characteristics of the audio signal.
For example, the audio label (``\textit{gunshots}'') indicates that the audio clip has sharp, high-decibel, and loud pops.
Therefore, acoustic labels provide explicit guidance for the audio clip contents and improve the generalization performance.
In addition to soft prompts, we provide additional explicit acoustic-aware prompts for decoding.

\vspace{2pt}
\noindent \textit{\textbf{Acoustic-aware Prompt:}} Firstly, we need to build the vocabulary of audio events $\mathcal{V}$.
We use the labels of AudioSet~\cite{gemmeke2017audio}, a prevalent benchmark dataset for the audio tagging task.
AudioSet contains 527 audio categories and covers various human and animal sounds, musical instruments and genres, and environmental sounds.
{ {Therefore, the audio events vocabulary $\mathcal{V}$ is a set of $527$ audio event labels $\left\{v_1,\dots,v_{527}\right\}$, where $v$ represents the audio event category.}}

Given the text embedding $f_{{clap}}^{{Text}}(T)$, we retrieve $M$ audio events that are most similar to $f_{{clap}}^{{Text}}(T)$ from the vocabulary $\mathcal{V}$ based on the cosine similarity of CLAP embeddings:
\begin{equation}
\left\{ v^{*}_{1},\dots, v^{*}_{M}\right\}= \left\{ \underset{v^{*}_m \in \mathcal{V}}{\mathrm{argmax}_{M}}\frac{f_{{clap}}^{{Text}}(T) \cdot f_{{clap}}^{{Text}}(v^{*}_m)}{\Vert f_{{clap}}^{{Text}}(T) \Vert\cdot\Vert f_{{clap}}^{{Text}}(v^{*}_m) \Vert}\right\},
\end{equation}
where $v^{*}_m$ is the $m$-th audio event. Therefore, the retrieved audio events are used to construct the hard prompt $H=$ ``There are $\left\{ v^{*}_{1},\dots, v^{*}_{M}\right\}$ in the audio.''.

We concatenate the hard prompts $H$ and the soft prompts $S$ along the sequence and feed them into the language decoder to reconstruct the input original text $T$ in an auto-regressive manner.
The model is trained using the cross-entropy loss:
\begin{equation}
\mathcal{L} = -\frac{1}{\left|T\right|}\sum_{i=1}^{\left|T\right|} \log p_{\theta}(t_i|T_{\textless i},H,S)
\end{equation}
where $\left|T\right|$ is the length of input $T$, $t_i$ is  $i$-th word  token of $T$, $T_{\textless i}$ includes all tokens from the start of 
$T$ up to just before the 
$i$-th token.
$p_{\theta}(\cdot)$ is the distribution of the output token and $\theta$ represents all parameters of the model.

\vspace{2pt}
\noindent \textit{\textbf{Prompt Dropout:}}
{To make the model robust to retrieval errors and diminish the effect of the modality gap in retrieval, we propose a simple but effective prompt dropout strategy, in which we randomly drop some audio categories in the hard prompts with dropout rate $\beta$ during training.
In this way, the model is trained to avoid simply concatenating audio events from hard prompts $H$ to generate the caption while ignoring the information in soft prompts $S$.}

\vspace{2pt}
\noindent \textit{\textbf{Zero-shot Inference:}}
For an input audio clip $A$, we use the CLAP audio encoder to replace the text encoder for extracting its audio embedding $f_{clap}^{Audio}(A)$. Following Kouzelis \emph{et al.}~\cite{kouzelis2023weakly}, we process the embedding $f_{clap}^{Audio}(A)$ in a similar way to get its soft prompts and hard prompts, excluding the mixed-augmentation and the prompt dropout strategy.
Next concatenated prompts are fed into the language decoder auto-regressively to generate the predicted descriptive caption $T$.

\section{Experimental Settings}

This section introduces the experimental settings, including model architectures, datasets, baselines and metrics, and implementation details.

\subsection{Model Architectures}
\vspace{2pt}
\noindent \textit{\textbf{CLAP Encoder:}} In this work, we use the CLAP model\footnote{\url{https://drive.google.com/drive/folders/1MeTBren6LaLWiZI8_phZvHvzz4r9QeCD}} as our encoder which is only trained on WavCaps~\cite{mei2023wavcaps}, which does not contain any human-annotated data.
The CLAP audio encoder is an \texttt{HTSAT}~\cite{chen2022hts} and the text encoder is a \texttt{RoBERTa}~\cite{liu2019roberta}.
All audio clips are randomly cropped or padded to $10$ seconds and sampled at a $32$k sampling rate. We use a $64$-dimensional log-Mel spectrogram extracted from a $1024$ point Hanning window with a hop size of $320$ as the input audio feature.
The dimension of the CLAP embedding is $1024$, and all parameters in the CLAP encoder are frozen.

\vspace{2pt}
\noindent \textit{\textbf{Mapping Network and Language Decoder:}}
The mapping network transforms the CLAP embedding $f_{clap}(\cdot)$ into soft prompts $S$.
This work employs a simple but effective mapping network containing only two linear layers.
For the language decoder, we use the pre-trained \texttt{GPT2-base}\footnote{\url{https://huggingface.co/openai-community/gpt2}}~\cite{radford2019language} to generate  text.
The dimension of hidden states is $768$, and all model parameters except the CLAP encoder are trainable. 

\subsection{Datasets}
We conduct our experiments on audio captioning benchmark datasets, AudioCaps~\cite{kim2019audiocaps} and Clotho~\cite{drossos2020clotho}.
AudioCaps is the largest human-annotated audio captioning dataset and contains $51$K audio clips with one caption per audio clip in the training set and five captions per audio clip in the evaluation set. {People annotated audio clips with the aid of visual information.}
Clotho is the official benchmark in the DCASE challenge.
{ Clotho contains about $3.8$K audio clips and each audio clip has five captions.}
The annotator uses the audio signals only for annotation and no additional signal is provided.

\subsection{Baselines}

\vspace{2pt}
\noindent \textit{\textbf{Fully Supervised Audio Captioning:}} 
We compare our method with fully supervised audio captioning methods: \textit{ACT}~\cite{mei2021audio}, \textit{MAAC}~\cite{ye2021improving}, \textit{{Xu~\textit{et al.}}}~\cite{xu2022sjtu}, \textit{Prefix AAC}~\cite{kim2023prefix}, \textit{RLSSR}~\cite{koh2022automated}, \textit{RECAP}~\cite{ghosh2024recap}, and \textit{ACTUAL}~\cite{zhang2023actual}.
{All of which are open source and not trained with additional data.}

\begin{table*}[]
\caption{Experimental results for in-domain scenarios on AudioCaps.}
\begin{threeparttable}
\begin{adjustbox}{width=0.9\linewidth,center}
\begin{tabular}{cccccccc}
\toprule[1.5pt]
{Method}         & BLEU$_1$ & BLEU$_4$ & ROUGE$_L$ & CIDEr & METEOR & SPICE & SPIDEr \\ \hline
\multicolumn{8}{c}{\mycc \textit{Fully Supervised Audio Captioning}}                                        \\  \hline

{Prefix AAC}~\cite{kim2023prefix} $\dagger$      & 71.3  & 30.9  & 50.3 & 73.3 & 24.0 & 17.7 & 45.5  \\

{RECAP}~
\cite{ghosh2024recap} $\dagger$           & \textbf{72.8}  & \textbf{31.7}  & \textbf{52.1} & \textbf{75.0} & \textbf{25.2}  & $18.3$ & \textbf{47.2}  \\ 
{ACT}~\cite{mei2021audio}           & 68.4 $\pm$ 0.44       & 25.2 $\pm$ 0.99       & 48.0 $\pm$ 0.35      &  67.5 $\pm$ 1.90          &  22.8 $\pm$ 0.27 & 16.9 $\pm$ 0.51      & 42.2 $\pm$  1.09\\
{MAAC}~\cite{ye2021improving}   & 64.0 $\pm$ 0.60  & 24.3 $\pm$ 0.55  & 44.7 $\pm$ 0.25  & 59.3 $\pm$ 1.05  & 21.0 $\pm$ 0.15  & 14.4 $\pm$ 0.38  & 36.9 $\pm$ 0.54  \\
{Xu~\textit{et al.}}~\cite{xu2022sjtu}    & 67.6 $\pm$ 0.21  & 27.2 $\pm$ 0.33  & 49.7 $\pm$ 0.17  & 73.8 $\pm$ 1.21  & 24.7 $\pm$ 0.06  & \textbf{18.4} $\pm$ 0.06  & 46.1 $\pm$ 0.62  \\\hline
\multicolumn{8}{c}{\mycc \textit{Zero-Shot Audio Captioning}}                                                   \\  \hline
{Audio Flamingo}~\cite{kong2024audio} $\dagger$  & $-$      & $-$     & $-$    & $50.2$ & $-$      & $-$    & $-$      \\
{Shaharabany~\textit{et al.}}~\cite{shaharabany2023zero} $\dagger$  & $-$   & 9.8  & 8.2 & 9.2 &  8.6 & $-$ & $-$ \\
{ZerAuCap}~\cite{salewski2023zero} $\dagger$       & $-$     & 6.8  & $33.1$ & $28.1$ & $12.3$  & $8.6$ & $18.3$  \\
{NoAudioCaptioning}~\cite{deshmukh2024training} & 59.2 $\pm$ 1.43        & 15.0 $\pm$ 0.66       & 40.4 $\pm$ 0.37        & 42.4 $\pm$ 1.58       & 19.6 $\pm$ 0.69       & 13.6 $\pm$ 0.51        & 28.0 $\pm$ 0.96       \\
{WSAC}~\cite{kouzelis2023weakly}           & 61.1 $\pm$ 0.48 & 17.1 $\pm$ 0.28 & 43.5 $\pm$ 0.36 & 56.4 $\pm$ 0.44 & \textbf{23.2} $\pm$ 0.09 & \textbf{16.3} $\pm$ 0.29 & 36.3 $\pm$ 0.31 \\ 
{Ours}               & \textbf{66.0} $\pm$ 0.15& \textbf{21.3} $\pm$ 0.48 & \textbf{45.7} $\pm$ 0.18 & \textbf{64.4} $\pm$ 0.61 & 22.0 $\pm$ 0.23 & 15.6 $\pm$ 0.23 & \textbf{40.0} $\pm$ 0.33 \\ \bottomrule[1.5pt]
\end{tabular}
\end{adjustbox}
\label{tab:indomain_audiocaps}
\begin{tablenotes}
   \footnotesize
   \item[$\dagger$] We use the original results listed in the paper since these works include results for in-domain and cross-domain scenarios.
 \end{tablenotes}
\end{threeparttable}
\vspace{-10pt}
\end{table*}
\begin{table*}[ht]

\caption{The experimental results for in-domain scenarios on the Clotho dataset}
\begin{threeparttable}
\begin{adjustbox}{width=0.9\linewidth,center}
\begin{tabular}{cccccccc}
\toprule[1.5pt]
{Method}            & BLEU$_1$  & BLEU$_4$  & ROUGE$_L$ & CIDEr   & METEOR & SPICE   & SPIDEr  \\  \hline
\multicolumn{8}{c}{\mycc \textit{Fully Supervised Audio Captioning} }                                                   \\  \hline

{Prefix AAC}~\cite{kim2023prefix}   $\dagger$       & 56.0   & 16.0   & 37.8 & 39.2   & 17.0  & 11.8   & 25.5   \\

{RECAP}~
\cite{ghosh2024recap}  $\dagger$             & 56.3   & 16.5   & {38.3} & {39.8}   & 17.9  & 12.2   & 21.4   \\
{ACTUAL}~\cite{zhang2023actual}  $\dagger$            & 56.6    & 16.1    & 37.5  & 40.9    & 17.6   & 12.1    & 26.5    \\
{RLSSR}~\cite{koh2022automated}  $\dagger$             & 55.1   & {16.8}   & 37.3 & 38.0    & 16.5  & 11.1   & 24.6   \\ 
{ACT}~\cite{mei2021audio}              & \textbf{58.4} $\pm$  0.21        & \textbf{16.9} $\pm$ 0.30        & \textbf{38.5}   $\pm$     0.30   &  41.6   $\pm$  0.46     & 17.8  $\pm$   0.08    & 12.1    $\pm$  0.14       &  26.9  $\pm$   0.26     \\
{MAAC}~\cite{ye2021improving} & {57.0} $\pm$ 0.53 & 16.0 $\pm$ 0.40 & 37.7 $\pm$ 0.37 & 41.3 $\pm$ 0.56& 17.7 $\pm$ 0.22& 12.3 $\pm$ 0.13& 26.8 $\pm$ 0.32\\
{Xu~\textit{et al.}}~\cite{xu2022sjtu}  & 56.9 $\pm$ 0.15& 16.0 $\pm$ 0.39& 37.9 $\pm$ 0.33 & \textbf{41.8} $\pm$ 0.69 & \textbf{17.9} $\pm$ 0.15& \textbf{12.7} $\pm$ 0.07 & \textbf{27.3} $\pm$ 0.34\\ \hline
\multicolumn{8}{c}{\mycc \textit{Zero-Shot Audio Captioning}}                                                            \\  \hline
{ZerAuCap}~\cite{salewski2023zero} $\dagger$          &   $-$ & 2.9     & 25.4  & 14.0    & 9.4    & 5.3     & 9.7     \\
{NoAudioCaptioning}~\cite{deshmukh2024training} & 51.8 $\pm$ 1.02  & 11.3 $\pm$ 0.80   & 34.7 $\pm$ 0.87 & 29.2 $\pm$ 1.25  & 15.6 $\pm$ 0.38 & 10.3 $\pm$ 0.24& 19.7 $\pm$ 0.66  \\
{WSAC}~\cite{kouzelis2023weakly}              & 54.5 $\pm$ 0.05 & 12.6 $\pm$ 0.14 & 35.9 $\pm$ 0.04& 35.7 $\pm$ 0.33 & 16.9 $\pm$ 0.02 & 11.8 $\pm$ 0.01 & 23.8 $\pm$ 0.17 \\
{Ours}               & \textbf{56.4} $\pm$ 0.24        & \textbf{15.6} $\pm$ 0.22 & \textbf{37.5} $\pm$ 0.17      &   \textbf{40.3} $\pm$ 0.47& \textbf{17.3} $\pm$ 0.17 & \textbf{11.9} $\pm$ 0.19 &  \textbf{26.1} $\pm$ 0.27       \\ \bottomrule[1.5pt]
\end{tabular}
\label{tab:indomain_clotho}
\end{adjustbox}
\begin{tablenotes}
   \footnotesize
   \item[$\dagger$] We use the original results listed in the paper since these works include results for in-domain and cross-domain scenarios.
 \end{tablenotes}
\end{threeparttable}
\vspace{-10pt}
\end{table*}

\vspace{2pt}
\noindent \textit{\textbf{Zero-Shot Audio Captioning:}}
We further compare our method with zero-shot audio captioning methods: \textit{Audio Flamingo}~\cite{kong2024audio}, \textit{Shaharabany et al.}~\cite{shaharabany2023zero}, \textit{ZerAuCap}~\cite{salewski2023zero}, \textit{NoAudioCaptioning}~\cite{deshmukh2024training}, and \textit{WSAC}~\cite{kouzelis2023weakly}.
Audio Flamingo~\cite{kong2024audio} is a large audio language model and achieves SOTA in several audio understanding tasks. Shaharabany et al.~\cite{shaharabany2023zero} and ZerAuCap~\cite{salewski2023zero} are decoder-guided zero-shot audio captioning methods.
NoAudioCaptioning~\cite{deshmukh2024training} and WSAC~\cite{kouzelis2023weakly}
are encoder-guided zero-shot audio captioning methods.

\subsection{Metrics}
Similar to other audio captioning works, we use common captioning metrics, including \textit{BLEU$_{n}$}~\cite{papineni2002bleu}, \textit{ROUGE$_L$}~\cite{lin2004rouge}, \textit{METEOR}~\cite{banerjee2005meteor}, \textit{CIDEr}~\cite{vedantam2015cider}, \textit{SPICE}~\cite{anderson2016spice}, and \textit{SPIDEr}~\cite{liu2017improved} for evaluation.
For all metrics, higher scores indicate better performance.

\subsection{Implementation Details}
In our work, we train the network using the AdamW optimizer with a weight decay of $0.02$, an initial learning rate of $1\times 10^{-5}$, a batch size of 32, a warm-up iteration of 3000 and a total training iteration of $15000$.
The model is trained on a $2080$Ti GPU.
We construct the hyperparameter tuning experiments and set $N=5$, $M=4$, $\sigma = 0.1$, $K=10$, and $\beta = 0.6$ for both AudioCaps and Clotho dataset.
We use beam search with a beam size of $3$ to generate captions during inference.

\section{Results and Discussion}
This section shows results followed by discussions of comparative experiments.
In all tables, the \textbf{bold} font represents the \textbf{best} result for each metric in the same setting.
Some works do not provide cross-domain results so we re-train these models using five different random seeds and report the mean and standard deviation of metrics.
\subsection{In-domain Audio Captioning}

Tables~\ref{tab:indomain_audiocaps} and~\ref{tab:indomain_clotho} compare our proposed method and baselines for in-domain scenarios, where the training and test sets come from the same benchmark dataset.
It should be specially noted that the zero-shot methods only use textual data from the training set for training, while the fully supervised methods use the audio-text paired data.
To make a fair comparison, we re-implement baseline zero-shot audio captioning methods using the same CLAP.

We have the following observations from the results for in-domain scenarios in the Clotho and AudioCaps datasets: 1) The fully supervised audio captioning methods tend to achieve better experimental performance than the zero-shot audio captioning methods. This is expected as the fully supervised methods are trained using audio-text pairs, and the models learn the ``audio-to-text" conversion ability well. The zero-shot methods suffer from the need to migrate from ``text-to-text'' in training to ``audio-to-text'' in inference, thus the discrepancy between training and inference results in worse in-domain performance.
2) Our proposed method outperforms other zero-shot audio captioning methods in most metrics.
We attribute this to the use of mixed augmentations and acoustic-aware prompts in model training, thereby mitigating the modality gap and improving the model's in-domain performance. 
3) Our proposed method, which does not utilize any paired data, achieves 86\% of the performance of the fully-supervised state-of-the-art method RECAP~\cite{ghosh2024recap}, which obtains a \textit{CIDEr} score of 75.0 on the AudioCaps dataset, and 95\% of the performance of the performance of Xu \textit{et al.}~\cite{xu2022sjtu}, which attains a \textit{CIDEr} score 41.8. This proves the effectiveness and practicality of our method.

\begin{table*}[ht]
\caption{The experimental results for Cross-domain scenarios on the AudioCaps and Clotho dataset}
\centering
\label{tab:crossdomain_ac_clotho}
\begin{threeparttable}
\begin{adjustbox}{width=0.9\linewidth,center}
\begin{tabular}{ccccccccc}
\toprule[1.5pt]
\multirow{2}{*}{Method} & \multicolumn{4}{c}{AudioCaps $\Longrightarrow$ Clotho}                           & \multicolumn{4}{c}{Clotho $\Longrightarrow$ AudioCaps}                       \\
\cmidrule(lr){2-5} \cmidrule(lr){6-9}
           & ROUGE$_L$ & CIDEr   & METEOR  & SPICE   & ROUGE$_L$ & CIDEr & METEOR  & SPICE  \\ \hline
\multicolumn{9}{c}{\mycc \textit{Fully Supervised Audio Captioning}}                                 \\ \hline

Prefix AAC~\cite{kim2023prefix} $\dagger$  & 27.6     & 19.2   & 11.2   & 7.4   & 33.0     & 21.1 & 14.4   & 8.3  \\
RECAP~
\cite{ghosh2024recap} $\dagger$       & 27.6     & 19.5   & 11.0   & 8.4   & 28.1     & 19.1 & 11.2   & 13.6  \\
ACT~\cite{mei2021audio}        & 26.1  $\pm$  0.44          &13.4   $\pm$   0.68     & 10.2 $\pm$ 0.25            & 5.5 $\pm$ 0.39    &  35.2 $\pm$    0.22      & 23.7  $\pm$   0.87   &  16.4   $\pm$    0.17      & 10.7   $\pm$    0.31    \\ 
MAAC~\cite{ye2021improving}         & 24.8 $\pm$ 0.83 & 16.4 $\pm$ 1.28& 10.3 $\pm$ 0.35 & 5.8 $\pm$ 0.10 & \textbf{35.9} $\pm$ 0.20 & 25.4 $\pm$ 0.45 & 17.1 $\pm$ 0.23 & 10.9 $\pm$ 0.18 \\
{Xu~\textit{et al.}}~\cite{xu2022sjtu}          & \textbf{29.2} $\pm$ 0.04    & \textbf{22.8} $\pm$  0.51     & \textbf{12.8} $\pm$ 0.07      & \textbf{8.5} $\pm$ 0.22    & {35.8} $\pm$ 0.29 & \textbf{25.6} $\pm$ 0.85 & \textbf{16.7} $\pm$ 0.30 & \textbf{11.1} $\pm$ 0.20 \\\hline

\multicolumn{9}{c}{\mycc \textit{Zero-shot Audio Captioning} }                                            \\  \hline
NoAudioCaptioning~\cite{deshmukh2024training}       & 26.6 $\pm$ 0.45  & 17.5 $\pm$ 2.00        & 11.1 $\pm$  0.59         & 7.4 $\pm$ 0.60      & 34.1 $\pm$  1.18        & 23.3  $\pm$ 1.68      & 16.7 $\pm$   0.36     & 10.6 $\pm$  0.34     \\
WSAC~\cite{kouzelis2023weakly}       & 26.6   $\pm$ 0.34 & 20.6 $\pm$ 0.31& 12.0 $\pm$ 0.11 & 8.2 $\pm$ 0.08& 35.5 $\pm$ 0.15   & 25.6 $\pm$ 0.22 & 17.3 $\pm$ 0.10 & 12.0 $\pm$ 
 0.08\\
Ours        & \textbf{29.8} $\pm$ 0.55  &  \textbf{24.8} $\pm$ 0.55       & \textbf{13.2} $\pm$ 0.46        & \textbf{9.3} $\pm$ 0.44        & \textbf{36.1} $\pm$ 0.51          &\textbf{33.8} $\pm$ 0.93& \textbf{18.0} $\pm$ 0.28        &\textbf{12.3} $\pm$ 0.18       \\ \bottomrule[1.5pt]
\end{tabular}
\end{adjustbox}
\begin{tablenotes}
   \footnotesize
   \item[$\dagger$] We use the original results listed in the paper since these works include results for in-domain and cross-domain scenarios.
 \end{tablenotes}
\end{threeparttable}
\vspace{-10pt}
\end{table*}

\subsection{Cross-domain Audio Captioning}
Cross-domain scenarios are where the training and test sets come from different benchmark datasets. The model is trained using only data from the \textit{Source} benchmark, and any data from the training set of the \textit{Target} benchmark is prohibited.
In the real world, the audio in \textit{Target} domain is often agnostic, so the cross-domain performance can better represent the effectiveness of the model in real-world applications.

Table~\ref{tab:crossdomain_ac_clotho} shows the experimental results of ours and baseline methods in cross-domain scenarios, where the ``$\textit{Source} \Longrightarrow \textit{Target}$'' refers to the scenario where the model is trained on the training set of the $\textit{Source}$ dataset and evaluated on the test set of the $\textit{Target}$ dataset.
It is important to note that neither the training nor the validation set of the $\textit{Target}$ dataset is used in model training and selection.
From the experimental results, we find the following: 1) Both fully supervised and zero-shot methods show some degree of degradation in the cross-domain scenarios compared to the in-domain scenarios. 2) Interestingly, the fully supervised methods do not exhibit significant superiority and achieve comparable results to zero-shot methods. We speculate that this might be because the strategies in the zero-shot methods help address the gap, improve model generalization, and reduce the risk of model over-fitting. 3) Our proposed model outperforms baselines across all metrics, including both fully supervised and zero-shot methods.
\begin{table*}[ht]

\caption{The experimental results under textual data from different fields}
\label{tab:crossdomain_different_source}
\centering
\begin{adjustbox}{width=0.95\linewidth,center}
\begin{tabular}{cccccccccc}

\toprule[1.5pt]
\multirow{2}{*}{Source Dataset} & \multirow{2}{*}{Size}&\multicolumn{4}{c}{Source Dataset $\Longrightarrow$ Clotho}                           & \multicolumn{4}{c}{Source Dataset $\Longrightarrow$ AudioCaps}                       \\\cmidrule(lr){3-6} \cmidrule(lr){7-10}
& & \multicolumn{1}{c}{ROUGE$_L$} & \multicolumn{1}{c}{CIDEr} & \multicolumn{1}{c}{METEOR} & \multicolumn{1}{c}{SPICE} & \multicolumn{1}{c}{ROUGE$_L$} & \multicolumn{1}{c}{CIDEr} & \multicolumn{1}{c}{METEOR} & \multicolumn{1}{c}{SPICE}\\ \hline
ChatGPT$^{\ref{footnote_chatgpt}}$ &31K  &25.5  $\pm$ 0.31  & 16.3 $\pm$ 0.62&10.6  $\pm$ 0.19  & 6.3  $\pm$ 0.11 &27.3  $\pm$ 0.27&15.5  $\pm$ 0.39&11.7  $\pm$ 0.17&7.1  $\pm$ 0.22\\
Freesound$^{\ref{footnote_fsd}}$ &84K & 30.4  $\pm$ 0.20 & 22.0 $\pm$ 0.84 &\textbf{12.6}  $\pm$ 0.20 &7.8  $\pm$ 0.15 &28.6  $\pm$ 0.42 &22.3  $\pm$ 0.94&12.3  $\pm$ 0.34& 6.7  $\pm$ 0.21\\
WavCaps~\cite{mei2023wavcaps} & 190K &\textbf{30.6} $\pm$ 0.36 &\textbf{22.1} $\pm$ 0.86  & \textbf{12.6} $\pm$ 0.22  & \textbf{7.9} $\pm$ 0.20 & \textbf{33.4} $\pm$ 1.21& \textbf{31.6} $\pm$ 1.96& \textbf{15.5} $\pm$ 0.61&\textbf{9.1} $\pm$ 0.59\\
COCO Captions~\cite{chen2015microsoft}      & 414K&25.9 $\pm$ 0.24  &10.0 $\pm$ 0.55& 8.9 $\pm$ 0.28 &5.1 $\pm$ 0.44 & 27.8 $\pm$ 0.53 &10.6 $\pm$ 1.11 &10.6 $\pm$ 0.55&6.2 $\pm$ 0.69\\

MusicCaps~\cite{agostinelli2023musiclm}    & 13K& 21.1  $\pm$ 1.34&6.6  $\pm$ 0.90& 8.8  $\pm$ 0.19&4.5  $\pm$ 0.42 &20.4  $\pm$ 1.84 &9.6  $\pm$ 0.42 &9.8  $\pm$ 0.13 &6.3  $\pm$ 0.89\\
LP-MusicCaps MSD~\cite{doh2023lp}     & 526K& 15.9 $\pm$ 0.72 &0.9 $\pm$ 0.10  &6.1 $\pm$ 0.11  &1.0 $\pm$ 0.16& 15.0 $\pm$ 0.56 & 0.8 $\pm$ 0.08 & 6.2 $\pm$ 0.24 & 0.9 $\pm$ 0.13\\ \bottomrule[1.5pt]
\end{tabular}
\end{adjustbox}
\end{table*}

Table~\ref{tab:crossdomain_different_source} shows the cross-domain performance of our proposed method trained on textual data from different fields and evaluated on Clotho and AudioCaps.
We use the textual data from three fields for training: audio captioning corpus (\textit{ChatGPT}\footnote{\url{https://chat.openai.com/}\label{footnote_chatgpt}}, \textit{FreeSound}\footnote{\url{https://freesound.org/}\label{footnote_fsd}}, \textit{WavCaps}~\cite{mei2023wavcaps}), visual captioning corpus (\textit{COCO Captions}~\cite{chen2015microsoft}), and music captioning corpus (\textit{MusicCaps}~\cite{agostinelli2023musiclm}, \textit{LP-MusicCaps MSD}~\cite{doh2023lp}).
For the text from \textit{ChatGPT}, we used GPT-3.5 to generate 31K text based on in-text learning. Specifically, we provide example captions from Clotho or AudioCaps and ask ChatGPT to generate similarly styled audio descriptions based on the examples.
The text data in \textit{FreeSound} comes from the subset of \textit{WavCaps}, collected through an online collaborative sound-sharing site.
\textit{WavCaps}~\cite{mei2023wavcaps} is a large-scale weakly-labeled audio captioning dataset that collects audio clips and their raw descriptions from web sources and uses ChatGPT to filter and clean noisy descriptions.
\textit{COCO Captions}~\cite{chen2015microsoft} is a human-annotated benchmark dataset in visual captioning.
For the music captioning corpus, \textit{MusicCaps}~\cite{agostinelli2023musiclm} is annotated by ten professional musicians and \textit{LP-MusicCaps MSD}~\cite{doh2023lp} is a large language model based pseudo music caption dataset.

From the results shown in Table~\ref{tab:crossdomain_different_source}, we have the following findings. 1) For the audio caption corpus generated by LLMs, the cross-domain performance on both Clotho and AudioCaps are improved by increasing the amount of textual data. 2) Compared to the results of the other methods shown in Table~\ref{tab:crossdomain_different_source}, our proposed method trained on weakly-labeled WavCaps achieves comparable cross-domain performance on Clotho and superior performance on AudioCaps, indicating the effectiveness of our proposed method. 3) The model trained on visual and music caption data exhibits worse cross-domain performance. This may be because CLAP is trained on weakly-labelled audio-caption paired data and cannot reconstruct the original caption from the other fields using its CLAP feature.

\begin{table*}[ht]
\caption{The ablation experiment results of different components.}
\label{tab:ablation}
\begin{adjustbox}{width=0.95\linewidth,center}

\begin{tabular}{ccccccccccccc}
\toprule[1.5pt]
\multicolumn{2}{c}{\multirow{2}{*}{Setting}} & \multicolumn{3}{c}{Components} & \multicolumn{4}{c}{In-Domain Scenarios} & \multicolumn{4}{c}{Cross-Domain Scenarios}                 \\ \cmidrule(lr){3-5} \cmidrule(lr){6-9} \cmidrule(lr){10-13} 
\multicolumn{2}{c}{}                         & IA       & EA       & AP       & ROUGE$_{L}$  & CIDEr & METEOR & SPICE & ROUGE$_{L}$  & CIDEr & METEOR & \multicolumn{1}{c}{SPICE} \\ \hline
\textit{Base Model} & \textit{a).}   &  &  & & 29.9  $\pm$ 0.82 &15.7  $\pm$ 0.37  &  13.1  $\pm$ 0.52& 7.5  $\pm$ 0.62    &29.8  $\pm$ 1.00& 13.8  $\pm$ 0.71 & 14.1  $\pm$ 0.62 & 7.8  $\pm$ 0.67\\ \hline
           & \textit{b).} & \checkmark &  &  & 33.0   $\pm$ 0.66 & 25.9  $\pm$ 0.97 & 15.0  $\pm$ 0.30 & 9.6  $\pm$ 0.34& 31.9  $\pm$ 0.28 & 18.2  $\pm$ 0.79 & 14.8  $\pm$ 0.30 &7.8  $\pm$ 0.49  \\
           & \textit{c).} & & \checkmark  &  &34.7 $\pm$ 0.30  &30.4 $\pm$ 0.89  & 15.7 $\pm$ 0.14 & 10.5 $\pm$ 0.32  &33.4 $\pm$ 0.21  &20.6 $\pm$ 0.48  &15.3 $\pm$ 0.05 &9.2 $\pm$ 0.15  \\
             & \textit{d).} &  &  & \checkmark & 35.0  $\pm$ 0.44 &  31.9  $\pm$ 0.93 &16.1  $\pm$ 0.17  &10.2  $\pm$ 0.18  &34.1  $\pm$ 0.52  &25.9  $\pm$ 0.88  & 16.6  $\pm$ 0.37 & 10.6  $\pm$ 0.45 \\
           & \textit{e).}  &\checkmark  & \checkmark  &  & 36.0 $\pm$ 0.27 &32.6 $\pm$ 0.46  & 16.1 $\pm$ 0.19 &11.0 $\pm$ 0.29  & 32.9 $\pm$ 0.35 &19.6 $\pm$ 0.76  &15.3 $\pm$ 0.19  &9.2 $\pm$ 0.17 \\
\textit{Full Model} &\textit{f).} &\checkmark  & \checkmark & \checkmark &   \textbf{37.5} $\pm$ 0.17 & \textbf{40.3} $\pm$ 0.47 &\textbf{17.3} $\pm$ 0.17   &\textbf{11.9} $\pm$ 0.19  & \textbf{36.1} $\pm$  0.51 & \textbf{33.8} $\pm$  0.93 &\textbf{18.0} $\pm$  0.28  & \textbf{12.3} $\pm$  0.18 \\ \bottomrule[1.5pt]
\end{tabular}
\end{adjustbox}
\vspace{-5pt}
\end{table*}
\subsection{Ablation Studies}
In this section, we conduct ablation experiments for in-domain and cross-domain scenarios by training the models on Clotho.
The results are shown in Table~\ref{tab:ablation}, where `IA', `EA', and `AP' are abbreviations for the instance augmentation, embedding augmentation, and acoustic-aware prompt, respectively.
The \textit{base model} does not use any components and its model structure only contains the CLAP encoder, the mapping network, and the language decoder.
The audio features are extracted using the CLAP audio encoder and fed into the trained mapping network and language decoder to generate the caption of the given audio during the inference stage. The model structure is shown in Fig.~\ref{fig:basemeth} (b).
The settings (\textit{b}, \textit{c}, \textit{d}) show that the components we proposed can improve the model performance in all metrics compared to the base model in setting \textit{a}.
In particular, the settings (\textit{b}, \textit{c}) show that both instance replacement and embedding augmentation can significantly improve the in-domain performance of the model.
These strategies reduce the modality gap between audio and text data, enhance the robustness of the model and improve the performance of zero-shot audio captioning.
Acoustic-aware prompts (setting \textit{d}) provide explicit guidance to the language decoder through hard prompts for audio events, thus enabling the model to achieve a better cross-domain generalization performance compared to the setting \textit{e}, with comparable in-domain performance.
Our \textit{full model} in the setting \textit{f} achieves significant improvements in all metrics (especially in the CIDEr metric) in both in-domain and cross-domain scenarios, indicating the effectiveness of our proposed model.

\subsection{Analysis on Hyper-parameters}

In the following, we conduct hyper-parameter tuning experiments to investigate and discuss the effects of different hyper-parameters on the model performance.
We fix the other hyper-parameters in the full model in each tuning experiment.

\subsubsection{The number of candidates $N$ in instance replacement}
\begin{table}[t]
\caption{The number of candidates $N$ in instance replacement}
\begin{adjustbox}{width=0.9\linewidth,center}
\begin{tabular}{ccccc}\toprule[1.5pt]
$N$  & ROUGE$_L$& CIDEr              & METEOR             & SPICE              \\ \hline
1 & 36.5 $\pm$ 0.17 & 35.8 $\pm$ 0.51 & 16.6 $\pm$ 0.11 & 11.2 $\pm$ 0.15\\
3  & 37.0 $\pm$ 0.15        & 36.8 $\pm$ 0.36     & \textbf{16.9} $\pm$ 0.10  &  \textbf{11.7} $\pm$ 0.18 \\
5   & \textbf{37.1} $\pm$ 0.10 & \textbf{36.9} $\pm$ 0.36 & \textbf{16.9} $\pm$ 0.10 & \textbf{11.7} $\pm$ 0.10\\
7  & 37.0 $\pm$ 0.18          & 36.6 $\pm$ 0.49 & 16.8 $\pm$ 0.09 &11.5 $\pm$ 0.14\\
10 & 37.2 $\pm$ 0.22          & 36.1 $\pm$ 0.70         & 16.8 $\pm$ 0.11  & 11.4 $\pm$ 0.13 \\ \bottomrule[1.5pt]
\end{tabular}
\label{tab:N}
\end{adjustbox}
\vspace{-15pt}
\end{table}
We first show the effect of the number of candidates $N$ in the instance replacement.
We select the number of candidates $N$ from values \{1, 3, 5, 7, 10\}.
The results are shown in Table~\ref{tab:N}. When $N$ is 5, the model performs better in most metrics.
As $N$ continues to increase, the model performance starts to deteriorate since augmented text samples contain texts that are far away from the original text for the model to learn an accurate ``text-to-text'' conversion. 

\subsubsection{The variance $\sigma^2$ of noise in embedding augmentation}
\begin{table}[t]
\caption{The variance $\sigma^2$ of noise in embedding augmentation}
\label{tab:variance}
\begin{adjustbox}{width=0.9\linewidth,center}
\begin{tabular}{ccccc}\toprule[1.5pt]
$\sigma^2$  & ROUGE$_L$     & CIDEr     & METEOR    & SPICE     \\ \midrule [1pt]
$1\times10^{-4}$ & 34.8 $\pm$ 0.72 & 31.0 $\pm$ 1.83 & 16.0 $\pm$ 0.43 & 10.1 $\pm$ 0.43 \\
$1\times10^{-3}$ & 35.4 $\pm$ 0.35 & 33.7 $\pm$ 0.58 & 16.3 $\pm$ 0.13 & 10.6 $\pm$ 0.18 \\
$1\times10^{-2}$ & \textbf{36.1} $\pm$ 0.31&\textbf{36.8} $\pm$ 0.45&\textbf{16.5} $\pm$ 0.09&\textbf{11.0} $\pm$ 0.12 \\
$1\times10^{-1}$ & 34.2 $\pm$ 0.17 & 32.1 $\pm$ 0.53 & 15.2 $\pm$ 0.11 & 9.7 $\pm$ 0.14 \\
$1$      & 34.4 $\pm$ 0.23 & 32.5 $\pm$ 0.75 &  15.3 $\pm$ 0.15 & 10.0 $\pm$ 0.21 \\ \bottomrule[1.5pt]
\end{tabular}
\end{adjustbox}
\vspace{-10pt}
\end{table}

In Table~\ref{tab:variance}, we present the results under different variances.
We find that the model performance is sensitive to the variance scale.
As the variance increases, the model performance improves progressively, suggesting that appropriate noise applied to the text embedding can significantly enhance the generalization ability of the model and weaken the effect of the modality gap. However, when the variance exceeds $1\times10^{-2}$, the model performance decreases rapidly due to excessive noise.

\subsubsection{The length $K$ of soft prompt}
We select the number of length $K$ from values \{1, 5, 10, 15, 20\}.
Table~\ref{tab:K} shows the experimental results under different lengths $K$.
We can find that the best performance is achieved in almost all metrics when $K$ is 10. {  When $K$ is 1, the inferior results are achieved because of the limited expressiveness of the model.}

\begin{table}[t]

\caption{The length $K$ of soft prompt}
\begin{adjustbox}{width=0.9\linewidth,center}
\label{tab:K}
\begin{tabular}{ccccc}\toprule[1.5pt]
$K$ & ROUGE$_L$     & CIDEr     & METEOR    & SPICE     \\ \midrule [1pt]
1     & 35.7 $\pm$ 0.29 & 35.5 $\pm$ 0.42 & 16.2 $\pm$ 0.11 & 10.5 $\pm$ 0.12\\
5     & 36.8 $\pm$ 0.04 & 39.2 $\pm$ 0.09 & 16.9 $\pm$ 0.04 & 11.4 $\pm$ 0.00\\
10    & \textbf{37.5} $\pm$ 0.17 &\textbf{40.3} $\pm$ 0.47   &\textbf{17.3} $\pm$ 0.17&\textbf{11.9} $\pm$ 0.19\\
15    & 37.2 $\pm$ 0.11 & 40.2 $\pm$ 0.78 & \textbf{17.3} $\pm$ 0.19 &11.6 $\pm$ 0.21 \\
20    & 37.1 $\pm$ 0.40 & 39.3 $\pm$ 2.40 & 17.2 $\pm$ 0.35 & 11.4 $\pm$ 0.12 \\ \bottomrule[1.5pt]
\end{tabular}
\end{adjustbox}
\vspace{-15pt}
\end{table}

\subsubsection{The number of audio events $M$ in hard prompt}

Table~\ref{tab:M} presents experimental results using different audio event numbers $M$.
The model performance is the best when we set $M$ to 4 or 5.
When $M$ is less than 4, the model performance improves with increasing $M$ due to more acoustic explicit information guidance.
However, when $M$ is greater than 5, the performance of the model decreases due to the increase in the irrelevance of the retrieved sound events. 
\begin{table}[t]

\caption{The number of audio events $M$ in the hard prompt}
\centering
\begin{adjustbox}{width=0.9\linewidth,center}
\begin{tabular}{ccccc}\toprule[1.5pt]
$M$ & ROUGE$_L$     & CIDEr     & METEOR    & SPICE     \\ \midrule [1pt]
1       & 33.8 $\pm$ 0.37 & 27.6 $\pm$ 0.98 & 15.0 $\pm$ 0.16 & 8.4 $\pm$ 1.70 \\
2       & 33.6 $\pm$ 0.64 &28.3 $\pm$ 0.47 &15.3 $\pm$ 0.22 &9.9 $\pm$ 0.15\\
3       & 33.9 $\pm$ 0.69 & 30.5 $\pm$ 1.56 & 15.7 $\pm$ 0.33 & 10.1 $\pm$ 0.38 \\
4       &  \textbf{35.3} $\pm$ 0.27 & \textbf{34.3} $\pm$ 0.71 & 16.1 $\pm$ 0.13 & 10.4 $\pm$ 0.21 \\
5       & 35.2 $\pm$ 0.29 & 34.1 $\pm$ 0.67 & \textbf{16.3} $\pm$ 0.18 & \textbf{10.5} $\pm$ 0.21  \\ 
7       & 34.8  $\pm$ 0.76 & 32.3  $\pm$ 1.22 & 15.8  $\pm$ 0.30	 & 10.3  $\pm$ 0.33\\ 
10       & 34.1  $\pm$ 0.20 & 31.5  $\pm$ 0.59	 & 15.3  $\pm$ 0.27	& 10.1  $\pm$ 0.14\\ 
\bottomrule[1.5pt]
\end{tabular}
\end{adjustbox}
\label{tab:M}
\vspace{-5pt}
\end{table}

\subsubsection{The Rate $\beta$ of prompt dropout}

\begin{table}[t]
\centering
\caption{The Rate $\beta$ of prompt dropout}
\begin{adjustbox}{width=0.9\linewidth,center}

\begin{tabular}{ccccc}\toprule[1.5pt]

$\beta$ & ROUGE$_L$      & CIDEr      & METEOR     & SPICE      \\ \midrule [1pt]
0     & 36.3  $\pm$ 0.64& 	34.7  $\pm$ 0.21 & 	16.9  $\pm$ 0.38 & 	11.4  $\pm$ 0.20 \\
0.2     & 36.7  $\pm$ 0.24 & 36.0  $\pm$ 0.68 & 17.1  $\pm$ 0.11	& 11.6  $\pm$ 0.25  \\
0.4     & 37.3 $\pm$ 0.12& 39.9 $\pm$ 0.21&17.2 $\pm$ 0.08&11.7 $\pm$ 0.18\\
0.6     & \textbf{37.5} $\pm$ 0.17 & \textbf{40.3} $\pm$ 0.47 & \textbf{17.3} $\pm$ 0.17 & \textbf{11.7} $\pm$ 0.19 \\
0.8     & 36.7  $\pm$ 0.56& 	37.5  $\pm$ 0.74&	17.2  $\pm$ 0.29	& 11.6  $\pm$ 0.42\\
1     &  36.1 $\pm$ 0.41&	35.4 $\pm$ 0.56& 16.3 $\pm$ 0.19& 11.2 $\pm$ 0.16\\ \bottomrule[1.5pt]
\end{tabular}
\end{adjustbox}
\label{tab:beta}
\vspace{-15pt}
\end{table}

Table~\ref{tab:beta} demonstrates the effect of different dropout rate $\beta$ on the performance.
We can see that the CIDEr score gradually increases as $\beta$ increases, indicating that dropout can prevent the model from relying heavily on the audio events and avoid the effects of retrieval errors and modality gaps.
When $\beta$ exceeds 0.6, the model performance decreases as useful audio events information is discarded so the model cannot leverage the explicit guidance.

\subsection{Multilingual Audio Captioning}
In addition, since only text is involved in the training stage, we can more easily use advanced language-based tools to investigate the potential applications of our proposed method, such as multilingual audio captioning, multi-styled audio captioning (literary style, children's style, etc.)

For example, when it comes to multilingual captioning systems, 
we use the Mistral~\cite{jiang2023mistral} large language model, which is a multilingual pre-trained text generation model with 7 billion parameters\footnote{\url{https://huggingface.co/mistralai/Mistral-7B-v0.1}}, to replace the GPT-2 as a language decoder for multilingual audio captioning.
We use the \textit{DeepL}\footnote{\url{https://www.deepl.com/}} to translate the Clotho English text data into different languages (Chinese, French).
{ The additional language token $L$ (e.g., $<$en$>$, $<$fr$>$) is fed into the language decoder with hard prompts $H$ and soft prompts $S$ to generate language-specific audio captions.}

\begin{table*}[t]
\centering
\caption{The in-domain experimental results on multilingual audio captioning}
\label{tab:multilingual}
\begin{adjustbox}{width=0.95\linewidth,center}
\begin{tabular}{cccccccccc}\toprule[1.5pt]
\multicolumn{1}{c}{\multirow{2}{*}{Setting}} & \multicolumn{3}{c}{English} & \multicolumn{3}{c}{French} & \multicolumn{3}{c}{Chinese} \\ \cmidrule(lr){2-4} \cmidrule(lr){5-7} \cmidrule(lr){8-10}
& {ROUGE$_L$} & {CIDEr} & {METEOR}  & {ROUGE$_L$} & {CIDEr} & {METEOR} & {ROUGE$_L$} & {CIDEr} & {METEOR} \\ \hline
Supervised Model &\textbf{37.9}  $\pm$ 0.28  &41.8  $\pm$ 1.31  &\textbf{17.6}  $\pm$ 0.21  &\textbf{29.8}  $\pm$ 2.81  & \textbf{29.3}  $\pm$ 2.76 &13.4  $\pm$ 1.22  &\textbf{28.2}  $\pm$ 0.94  &\textbf{20.6}  $\pm$ 1.58  & \textbf{14.6}  $\pm$ 0.33   \\
ZS-Base Model &32.3  $\pm$ 0.82 &24.4  $\pm$ 1.18  &14.3  $\pm$ 0.51 &26.0  $\pm$ 0.56  &18.5  $\pm$ 1.46  &12.4  $\pm$ 0.35  &25.5  $\pm$ 1.06  &15.7  $\pm$ 2.34  &13.9  $\pm$ 0.36   \\
ZS-Full Model &37.7  $\pm$ 0.44  &\textbf{42.1}  $\pm$ 0.50  &\textbf{17.6}  $\pm$ 0.13  &29.6  $\pm$ 2.84 &28.6  $\pm$ 3.25  & \textbf{13.5}  $\pm$ 1.29 &27.9  $\pm$ 0.52 &20.4  $\pm$ 0.34  & 14.3  $\pm$ 0.19  \\ \bottomrule[1.5pt]
\end{tabular}
\end{adjustbox}
\vspace{-5pt}
\end{table*}

The results are shown in Table~\ref{tab:multilingual}, where `ZS' is the abbreviation for zero-shot.
Our proposed method, the ZS-Full Model, achieves comparable results with the fully supervised method in most metrics and even achieves better results in English compared to the experimental results in Table~\ref{tab:indomain_clotho}.
We believe that Mistral has more powerful text generation capabilities compared to GPT-2, and therefore can exploit multimodal semantic information and generate descriptive text more accurately.
In addition, the ZS-Base Model still achieves inferior performance in all the metrics compared to our proposed method, the ZS-Full Model, which demonstrates that our proposed mixed-augmentation-based soft prompt strategy and the retrieval-based acoustic-aware hard prompt strategy can also improve the generalization performance of zero-shot audio captioning in the multi-lingual scenario. 
\subsection{Qualitative Analysis}
\subsubsection{In-domain Audio Captioning}
\begin{table*}[htbp]
\setlength{\tabcolsep}{1pt}
\caption{The sample results of the in-domain audio captioning}
\label{tab:in_sample}
\begin{adjustbox}{width=0.95\linewidth,center}
\begin{tabular}{c p{5.5cm}<{\centering}p{5.5cm}<{\centering}p{5.5cm}<{\centering}p{5.5cm}<{\centering}}
\toprule[1.5pt]
\multirow{2}{*}{Sample} & \multicolumn{2}{c}{AudioCaps} & \multicolumn{2}{c}{Clotho} \\ \cmidrule(lr){2-3} \cmidrule(lr){4-5}
 & YqeSl7YZAfs4.wav & YonBZOH88OYs.wav & t34t trafik{[}1{]}.wav & Ronda - The Old Shrine - La antigua Ermita.wav \\ \hline
Ground Truth & \textcolor[RGB]{179,0,0}{\textit{faucet}} \textcolor[RGB]{63,117,185}{\textit{running}} and a  \textcolor[RGB]{179,0,0}{\textit{man}} \textcolor[RGB]{63,117,185}{\textit{speaks}} & repeated bursts of \textcolor[RGB]{63,117,185}{\textit{spray}} & \textcolor[RGB]{179,0,0}{\textit{car horns}} \textcolor[RGB]{63,117,185}{\textit{honk}} in traffic and \textcolor[RGB]{179,0,0}{\textit{people}} \textcolor[RGB]{63,117,185}{\textit{shout}} in the background & \textcolor[RGB]{179,0,0}{\textit{birds}} are \textcolor[RGB]{63,117,185}{\textit{singing}} while \textcolor[RGB]{179,0,0}{\textit{people}} \textcolor[RGB]{63,117,185}{\textit{talk}} in the background \\
Prediction & a \textcolor[RGB]{179,0,0}{\textit{man}} is \textcolor[RGB]{63,117,185}{\textit{speaking}} and \textcolor[RGB]{179,0,0}{\textit{water}} is \textcolor[RGB]{63,117,185}{\textit{running}} from a faucet & \textcolor[RGB]{63,117,185}{\textit{spraying}} and \textcolor[RGB]{63,117,185}{\textit{hissing}} & \textcolor[RGB]{179,0,0}{\textit{cars}} are \textcolor[RGB]{63,117,185}{\textit{honking}} their horns and \textcolor[RGB]{179,0,0}{\textit{people}} are \textcolor[RGB]{63,117,185}{\textit{talking}} in the background & \textcolor[RGB]{179,0,0}{\textit{birds}} are \textcolor[RGB]{63,117,185}{\textit{chirping}} and \textcolor[RGB]{179,0,0}{\textit{people}} are \textcolor[RGB]{63,117,185}{\textit{talking}} in the background\\
Audio Events & water tap, faucet, sink (filling or washing), bathtub (filling or washing), male speech, man speaking & spray, hiss, air brake, steam & vehicle horn, car horn, honking, honk, air horn, truck horn, traffic noise, roadway noise & country, bird, field recording, noise \\\bottomrule[1.5pt]
\end{tabular}
\end{adjustbox}
\vspace{-5pt}
\end{table*} Table~\ref{tab:in_sample} shows the visualization results for the AudioCaps and Clotho datasets in the in-domain setting, where \textcolor[RGB]{179,0,0}{\textit{red}} and \textcolor[RGB]{63,117,185}{\textit{blue}} are {the sound events objects and their actions behavior, respectively}. The last row is the retrieved audio events in the acoustic-aware prompts.
We can find that benefiting from the explicit guidance provided by the acoustic-aware prompt and from the bridge to close the modality gap in the multimodal semantic space provided by the mixed-augmentation strategy,  our proposed zero-shot method does not use any paired audio-text data for training, but can still accurately recognize the audio events and describe the contents of the audio clip during inference.
In addition, the prompt dropout can mitigate the over-reliance of the model on explicit prompts: in the fourth sample, the retrieved sound events provide irrelevant information (`country', `field recording', and `noise'), but the model manages to generate accurate descriptions, overcoming the interference of noisy guidance.

\subsubsection{Cross-domain Audio Captioning}
\begin{table*}[htbp]
\setlength{\tabcolsep}{1pt}
\caption{The sample results of the cross-domain audio captioning}
\begin{adjustbox}{width=0.95\linewidth,center}
\begin{tabular}{c p{5.5cm}<{\centering}p{5.5cm}<{\centering}p{5.5cm}<{\centering}p{5.5cm}<{\centering}}
\toprule[1.5pt]
 \multicolumn{1}{c}{\multirow{2}{*}{Sample}}& Clotho $\Longrightarrow$ AudioCaps & AudioCaps $\Longrightarrow$ Clotho & ChatGPT $\Longrightarrow$ AudioCaps & WavCaps $\Longrightarrow$ Clotho \\ \cmidrule(lr){2-2} \cmidrule(lr){3-3} \cmidrule(lr){4-4} \cmidrule(lr){5-5}
 & YfBYDJWChe5c.wav & Blade Big.wav & YwoadpeAGHUQ.wav & steam train 05.wav \\ \hline
Ground Truth & a \textcolor[RGB]{179,0,0}{\textit{person}} \textcolor[RGB]{63,117,185}{\textcolor[RGB]{63,117,185}{\textit{snoring}}}& \textcolor[RGB]{179,0,0}{{\textit{metal}}} \textcolor[RGB]{63,117,185}{\textcolor[RGB]{63,117,185}{\textit{sliding}}} together such as \textcolor[RGB]{179,0,0}{{\textit{swords}}} or \textcolor[RGB]{179,0,0}{\textit{knives}}& an \textcolor[RGB]{179,0,0}{\textit{emergency siren}} \textcolor[RGB]{63,117,185}{\textit{blaring}} steadily & a \textcolor[RGB]{179,0,0}{\textit{person}} \textcolor[RGB]{63,117,185}{\textit{talks}} on board a \textcolor[RGB]{179,0,0}{\textit{train}} while it \textcolor[RGB]{63,117,185}{\textit{rattles}} along the tracks \\
Prediction & a \textcolor[RGB]{179,0,0}{\textit{person}} is \textcolor[RGB]{63,117,185}{\textit{snoring}} & \textcolor[RGB]{63,117,185}{\textit{clanking}} and \textcolor[RGB]{63,117,185}{\textit{clanking}} & an \textcolor[RGB]{179,0,0}{\textit{ambulance siren}} \textcolor[RGB]{63,117,185}{\textit{wails}} urgently, demanding attention from its passengers & a \textcolor[RGB]{179,0,0}{train} is \textcolor[RGB]{63,117,185}{\textit{moving}} on a track with a \textcolor[RGB]{63,117,185}{\textit{clickety-clack}} sound\\
Audio Events & snoring, snort, babbling, groan & dishes, pots, and pans, cutlery, silverware, scrape, heavy metal & fire engine, fire truck (siren), emergency vehicle, ambulance (siren) & train, railroad car, train wagon, clickety-clack \\ \bottomrule[1.5pt]
\end{tabular}
\label{tab:cross_sample}
\end{adjustbox}
\vspace{-10pt}
\end{table*}

We also present the ground truth captions and the generated captions of our proposed method in the cross-domain setting, shown in Table~\ref{tab:cross_sample}.
We can observe that the training corpus has a tremendous impact on the style of the generated text. For instance, in the second sample, the training set of AudioCaps contains lots of short, generalized text, which results in concise captions.
In the third sample, the text generated by ChatGPT results in speculative descriptions ``\textit{demanding attention from its
passengers}''.

\subsubsection{Multilingual Audio Captioning}
\begin{table}[]
\setlength{\tabcolsep}{1pt}
\caption{The sample results of the multilingual audio captioning}
\begin{adjustbox}{width=0.95\linewidth,center}
\begin{tabular}{cp{5cm}<{\centering}p{5cm}<{\centering}}\toprule[1.5pt]
Sample & enoesque-Thunder and Rain 1.wav & Pencil Writing.wav \\ \midrule
Ground Truth & \textcolor[RGB]{179,0,0}{\textit{rain}} starts \textcolor[RGB]{63,117,185}{\textit{pouring down}} and \textcolor[RGB]{179,0,0}{\textit{thunder}} \textcolor[RGB]{63,117,185}{\textit{makes}} a boom & a \textcolor[RGB]{179,0,0}{\textit{person}} \textcolor[RGB]{63,117,185}{\textit{writes}} several words on a chalkboard \\
English & \textcolor[RGB]{179,0,0}{\textit{thunder}} is  \textcolor[RGB]{63,117,185}{\textit{rumbling}} and \textcolor[RGB]{179,0,0}{\textit{rain}} is \textcolor[RGB]{63,117,185}{\textit{falling}} & a \textcolor[RGB]{179,0,0}{\textit{person}} is  \textcolor[RGB]{63,117,185}{\textit{writing}} on a chalkboard with chalk \\
French & la \textcolor[RGB]{179,0,0}{\textit{pluie}} \textcolor[RGB]{63,117,185}{\textit{tombe}} sur le sol à un rythme régul. & \textcolor[RGB]{179,0,0}{\textit{quelqu'un}} \textcolor[RGB]{63,117,185}{\textit{écrit}} sur un tableau \\
Chinese & \begin{CJK}{UTF8}{gbsn}{\textcolor[RGB]{179,0,0}{\textit{大雨}}\textcolor[RGB]{63,117,185}{\textit{倾盆而下}}}\end{CJK} & \begin{CJK}{UTF8}{gbsn}{有\textcolor[RGB]{179,0,0}{\textit{人}}在黑板上\textcolor[RGB]{63,117,185}{\textit{写}}字}\end{CJK} \\ \bottomrule[1.5pt]
\end{tabular}
\label{tab:multilingual_sample}
\end{adjustbox}
\vspace{-15pt}
\end{table}
Table~\ref{tab:multilingual_sample} shows the samples of English, French, and Chinese audio captions generated by our proposed model. Our method can generate descriptive text for the corresponding audio in an end-to-end process, regardless of the language, providing a solid basis for applying the multilingual audio captioning method.

\section{Conclusion and Feature Works}
We have presented a novel zero-shot audio captioning method that does not employ human-labeled audio-text paired data but only uses the text corpus for model training. Our proposed method avoids the reliance on highly costly paired data. To bridge the modality gap of multimodal semantic space and to enhance the generalization performance of the model, we devise a mixed-augmentation strategy and a retrieval-based acoustic-aware prompt strategy. Extensive experiments were conducted on AudioCaps and Clotho to demonstrate the effectiveness of our proposed method. Our proposed method performs better on most metrics for the in-domain setting than other zero-shot audio captioning methods. In the cross-domain setting, our proposed method outperforms the compared methods in all metrics, both fully supervised and zero-shot audio captioning methods.
Moreover, our proposed method shows the potential of multilingual audio captioning.
Experimental results show that our method can generate multilingual descriptive text for input audio in an end-to-end style.

For future work, we plan to explore the effectiveness of our proposed method in other audio-text multimodal tasks, such as Music Captioning and Audio Question Answering tasks.
Moreover, we plan to perform further research on multilingual and multi-styled audio captioning methods to promote the democratization of audio captioning.

\bibliographystyle{IEEEtran}
\bibliography{ref}

\begin{thebibliography}{10}
\providecommand{\url}[1]{#1}
\csname url@samestyle\endcsname
\providecommand{\newblock}{\relax}
\providecommand{\bibinfo}[2]{#2}
\providecommand{\BIBentrySTDinterwordspacing}{\spaceskip=0pt\relax}
\providecommand{\BIBentryALTinterwordstretchfactor}{4}
\providecommand{\BIBentryALTinterwordspacing}{\spaceskip=\fontdimen2\font plus
\BIBentryALTinterwordstretchfactor\fontdimen3\font minus \fontdimen4\font\relax}
\providecommand{\BIBforeignlanguage}[2]{{%
\expandafter\ifx\csname l@#1\endcsname\relax
\typeout{** WARNING: IEEEtran.bst: No hyphenation pattern has been}%
\typeout{** loaded for the language `#1'. Using the pattern for}%
\typeout{** the default language instead.}%
\else
\language=\csname l@#1\endcsname
\fi
#2}}
\providecommand{\BIBdecl}{\relax}
\BIBdecl

\bibitem{drossos2017automated}
K.~Drossos, S.~Adavanne, and T.~Virtanen, ``Automated audio captioning with recurrent neural networks,'' in \emph{2017 IEEE Workshop on Applications of Signal Processing to Audio and Acoustics (WASPAA)}, 2017, pp. 374--378.

\bibitem{drossos2020clotho}
K.~Drossos, S.~Lipping, and T.~Virtanen, ``Clotho: An audio captioning dataset,'' in \emph{IEEE International Conference on Acoustics, Speech and Signal Processing (ICASSP)}, 2020, pp. 736--740.

\bibitem{kim2019audiocaps}
C.~D. Kim, B.~Kim, H.~Lee, and G.~Kim, ``Audio\uppercase{C}aps: Generating captions for audios in the wild,'' in \emph{Proceedings of the 2019 Conference of the North American Chapter of the Association for Computational Linguistics: Human Language Technologies, Volume 1 (Long and Short Papers)}, 2019, pp. 119--132.

\bibitem{lipping2019crowdsourcing}
S.~Lipping, K.~Drossos, and T.~Virtanen, ``Crowdsourcing a dataset of audio captions,'' in \emph{Acoustic Scenes and Events 2019 Workshop (DCASE2019)}, 2019, p. 139.

\bibitem{xu2023beyond}
X.~Xu, Z.~Xie, M.~Wu, and K.~Yu, ``Beyond the status quo: A contemporary survey of advances and challenges in audio captioning,'' \emph{IEEE/ACM Transactions on Audio, Speech, and Language Processing}, 2023.

\bibitem{chen2015microsoft}
X.~Chen, H.~Fang, T.-Y. Lin, R.~Vedantam, S.~Gupta, P.~Doll{\'a}r, and C.~L. Zitnick, ``Microsoft \uppercase{coco} \uppercase{c}aptions: Data collection and evaluation server,'' \emph{arXiv preprint arXiv:1504.00325}, 2015.

\bibitem{zhang2023actual}
Y.~Zhang, H.~Yu, R.~Du, Z.-H. Tan, W.~Wang, Z.~Ma, and Y.~Dong, ``\uppercase{ACTUAL}: Audio captioning with caption feature space regularization,'' \emph{IEEE/ACM Transactions on Audio, Speech, and Language Processing}, 2023.

\bibitem{martin2021diversity}
I.~M. Morato and A.~Mesaros, ``Diversity and bias in audio captioning datasets,'' in \emph{Detection and Classication of Acoustic Scenes and Events}, 2021, pp. 90--94.

\bibitem{wu2023large}
Y.~Wu, K.~Chen, T.~Zhang, Y.~Hui, T.~Berg-Kirkpatrick, and S.~Dubnov, ``Large-scale contrastive language-audio pretraining with feature fusion and keyword-to-caption augmentation,'' in \emph{ICASSP 2023-2023 IEEE International Conference on Acoustics, Speech and Signal Processing (ICASSP)}.\hskip 1em plus 0.5em minus 0.4em\relax IEEE, 2023, pp. 1--5.

\bibitem{mei2023wavcaps}
X.~Mei, C.~Meng, H.~Liu, Q.~Kong, T.~Ko, C.~Zhao, M.~D. Plumbley, Y.~Zou, and W.~Wang, ``Wav\uppercase{c}aps: A chatgpt-assisted weakly-labelled audio captioning dataset for audio-language multimodal research,'' \emph{arXiv preprint arXiv:2303.17395}, 2023.

\bibitem{liang2022mind}
V.~W. Liang, Y.~Zhang, Y.~Kwon, S.~Yeung, and J.~Y. Zou, ``Mind the gap: Understanding the modality gap in multi-modal contrastive representation learning,'' \emph{Advances in Neural Information Processing Systems}, vol.~35, pp. 17\,612--17\,625, 2022.

\bibitem{10095889}
B.~Elizalde, S.~Deshmukh, M.~A. Ismail, and H.~Wang, ``Clap learning audio concepts from natural language supervision,'' in \emph{ICASSP 2023 - 2023 IEEE International Conference on Acoustics, Speech and Signal Processing (ICASSP)}, 2023, pp. 1--5.

\bibitem{elizalde2024natural}
B.~Elizalde, S.~Deshmukh, and H.~Wang, ``Natural language supervision for general-purpose audio representations,'' in \emph{ICASSP 2024-2024 IEEE International Conference on Acoustics, Speech and Signal Processing (ICASSP)}.\hskip 1em plus 0.5em minus 0.4em\relax IEEE, 2024, pp. 336--340.

\bibitem{kong2020panns}
Q.~Kong, Y.~Cao, T.~Iqbal, Y.~Wang, W.~Wang, and M.~D. Plumbley, ``\uppercase{Pann}s: Large-scale pretrained audio neural networks for audio pattern recognition,'' \emph{IEEE/ACM Transactions on Audio, Speech, and Language Processing}, vol.~28, pp. 2880--2894, 2020.

\bibitem{chen2022hts}
K.~Chen, X.~Du, B.~Zhu, Z.~Ma, T.~Berg-Kirkpatrick, and S.~Dubnov, ``\uppercase{HTS-AT}: A hierarchical token-semantic audio transformer for sound classification and detection,'' in \emph{ICASSP 2022-2022 IEEE International Conference on Acoustics, Speech and Signal Processing (ICASSP)}.\hskip 1em plus 0.5em minus 0.4em\relax IEEE, 2022, pp. 646--650.

\bibitem{devlin2018bert}
J.~Devlin, M.-W. Chang, K.~Lee, and K.~Toutanova, ``\uppercase{Bert}: Pre-training of deep bidirectional transformers for language understanding,'' \emph{arXiv preprint arXiv:1810.04805}, 2018.

\bibitem{liu2019roberta}
Y.~Liu, M.~Ott, N.~Goyal, J.~Du, M.~Joshi, D.~Chen, O.~Levy, M.~Lewis, L.~Zettlemoyer, and V.~Stoyanov, ``\uppercase{Roberta}: A robustly optimized bert pretraining approach,'' \emph{arXiv preprint arXiv:1907.11692}, 2019.

\bibitem{oord2018representation}
A.~v.~d. Oord, Y.~Li, and O.~Vinyals, ``Representation learning with contrastive predictive coding,'' \emph{arXiv preprint arXiv:1807.03748}, 2018.

\bibitem{mei2021audio}
X.~Mei, X.~Liu, Q.~Huang, M.~D. Plumbley, and W.~Wang, ``Audio captioning transformer,'' \emph{arXiv preprint arXiv:2107.09817}, 2021.

\bibitem{ye2021improving}
Z.~Ye, H.~Wang, D.~Yang, and Y.~Zou, ``Improving the performance of automated audio captioning via integrating the acoustic and semantic information,'' \emph{arXiv preprint arXiv:2110.06100}, 2021.

\bibitem{xu2022sjtu}
X.~Xu, Z.~Xie, M.~Wu, and K.~Yu, ``The \uppercase{SJTU} system for dcase2022 challenge task 6: Audio captioning with audio-text retrieval pre-training,'' \emph{DCASE 2022 Challenge, Tech. Rep.}, 2022.

\bibitem{kim2023prefix}
M.~Kim, K.~Sung-Bin, and T.-H. Oh, ``Prefix tuning for automated audio captioning,'' in \emph{IEEE International Conference on Acoustics, Speech and Signal Processing (ICASSP)}, 2023.

\bibitem{koh2022automated}
A.~Koh, X.~Fuzhao, and C.~E. Siong, ``Automated audio captioning using transfer learning and reconstruction latent space similarity regularization,'' in \emph{IEEE International Conference on Acoustics, Speech and Signal Processing (ICASSP)}, 2022, pp. 7722--7726.

\bibitem{ghosh2024recap}
S.~Ghosh, S.~Kumar, C.~K.~R. Evuru, R.~Duraiswami, and D.~Manocha, ``\uppercase{RECAP}: retrieval-augmented audio captioning,'' in \emph{ICASSP 2024-2024 IEEE International Conference on Acoustics, Speech and Signal Processing (ICASSP)}.\hskip 1em plus 0.5em minus 0.4em\relax IEEE, 2024, pp. 1161--1165.

\bibitem{shaharabany2023zero}
T.~Shaharabany, A.~Shaulov, and L.~Wolf, ``Zero-shot audio captioning via audibility guidance,'' \emph{arXiv preprint arXiv:2309.03884}, 2023.

\bibitem{kong2024audio}
Z.~Kong, A.~Goel, R.~Badlani, W.~Ping, R.~Valle, and B.~Catanzaro, ``Audio \uppercase{F}lamingo: A novel audio language model with few-shot learning and dialogue abilities,'' \emph{arXiv preprint arXiv:2402.01831}, 2024.

\bibitem{salewski2023zero}
L.~Salewski, S.~Fauth, A.~Koepke, and Z.~Akata, ``Zero-shot audio captioning with audio-language model guidance and audio context keywords,'' \emph{arXiv preprint arXiv:2311.08396}, 2023.

\bibitem{deshmukh2024training}
S.~Deshmukh, B.~Elizalde, D.~Emmanouilidou, B.~Raj, R.~Singh, and H.~Wang, ``Training audio captioning models without audio,'' in \emph{ICASSP 2024-2024 IEEE International Conference on Acoustics, Speech and Signal Processing (ICASSP)}.\hskip 1em plus 0.5em minus 0.4em\relax IEEE, 2024, pp. 371--375.

\bibitem{kouzelis2023weakly}
T.~Kouzelis and V.~Katsouros, ``Weakly-supervised automated audio captioning via text only training,'' \emph{arXiv preprint arXiv:2309.12242}, 2023.

\bibitem{gemmeke2017audio}
J.~F. Gemmeke, D.~P. Ellis, D.~Freedman, A.~Jansen, W.~Lawrence, R.~C. Moore, M.~Plakal, and M.~Ritter, ``Audio\uppercase{s}et: An ontology and human-labeled dataset for audio events,'' in \emph{2017 IEEE international conference on acoustics, speech and signal processing (ICASSP)}.\hskip 1em plus 0.5em minus 0.4em\relax IEEE, 2017, pp. 776--780.

\bibitem{radford2019language}
A.~Radford, J.~Wu, R.~Child, D.~Luan, D.~Amodei, I.~Sutskever \emph{et~al.}, ``Language models are unsupervised multitask learners,'' \emph{OpenAI blog}, vol.~1, no.~8, p.~9, 2019.

\bibitem{papineni2002bleu}
K.~Papineni, S.~Roukos, T.~Ward, and W.-J. Zhu, ``\uppercase{Bleu}: a method for automatic evaluation of machine translation,'' in \emph{Proceedings of the 40th annual meeting of the Association for Computational Linguistics}, 2002, pp. 311--318.

\bibitem{lin2004rouge}
C.-Y. Lin, ``\uppercase{Rouge}: A package for automatic evaluation of summaries,'' in \emph{Text summarization branches out}, 2004, pp. 74--81.

\bibitem{banerjee2005meteor}
S.~Banerjee and A.~Lavie, ``\uppercase{METEOR}: An automatic metric for mt evaluation with improved correlation with human judgments,'' in \emph{Proceedings of the acl workshop on intrinsic and extrinsic evaluation measures for machine translation and/or summarization}, 2005, pp. 65--72.

\bibitem{vedantam2015cider}
R.~Vedantam, C.~Lawrence~Zitnick, and D.~Parikh, ``\uppercase{Cider}: Consensus-based image description evaluation,'' in \emph{Proceedings of the IEEE conference on computer vision and pattern recognition}, 2015, pp. 4566--4575.

\bibitem{anderson2016spice}
P.~Anderson, B.~Fernando, M.~Johnson, and S.~Gould, ``\uppercase{Spice}: Semantic propositional image caption evaluation,'' in \emph{European conference on computer vision}.\hskip 1em plus 0.5em minus 0.4em\relax Springer, 2016, pp. 382--398.

\bibitem{liu2017improved}
S.~Liu, Z.~Zhu, N.~Ye, S.~Guadarrama, and K.~Murphy, ``Improved image captioning via policy gradient optimization of spider,'' in \emph{Proceedings of the IEEE international conference on computer vision}, 2017, pp. 873--881.

\bibitem{agostinelli2023musiclm}
A.~Agostinelli, T.~I. Denk, Z.~Borsos, J.~Engel, M.~Verzetti, A.~Caillon, Q.~Huang, A.~Jansen, A.~Roberts, M.~Tagliasacchi \emph{et~al.}, ``Music\uppercase{lm}: Generating music from text,'' \emph{arXiv preprint arXiv:2301.11325}, 2023.

\bibitem{doh2023lp}
S.~Doh, K.~Choi, J.~Lee, and J.~Nam, ``\uppercase{LP}-\uppercase{M}usic\uppercase{C}aps: Llm-based pseudo music captioning,'' \emph{arXiv preprint arXiv:2307.16372}, 2023.

\bibitem{jiang2023mistral}
A.~Q. Jiang, A.~Sablayrolles, A.~Mensch, C.~Bamford, D.~S. Chaplot, D.~d.~l. Casas, F.~Bressand, G.~Lengyel, G.~Lample, L.~Saulnier \emph{et~al.}, ``Mistral 7b,'' \emph{arXiv preprint arXiv:2310.06825}, 2023.

\end{thebibliography}
\end{document}